\documentclass[%
reprint,
superscriptaddress,
 amsmath,amssymb,
 aps,
prb,
]{revtex4-2}

\usepackage{graphicx}
\usepackage{dcolumn}
\usepackage{bm}

\newcommand{\vo}{V\textsubscript{2}O\textsubscript{3}} 
\newcommand{\sap}{Al\textsubscript{2}O\textsubscript{3} } 
\newcommand{\is}{\textit{in-situ} } 
\newcommand{\es}{\textit{ex-situ} } 

\begin{document}

\title{Metal to insulator transition at the surface of V\textsubscript{2}O\textsubscript{3} thin films: an \is view}
\author{M. Caputo}
\affiliation{Swiss Light Source, Paul Scherrer Institut, CH-5232 Villigen, Switzerland}
\affiliation{Elettra Sincrotrone Trieste, s.s. 14 km 163.5 in Area Science Park, 34149 Trieste, Italy}
\author{J. Jandke}
\affiliation{Swiss Light Source, Paul Scherrer Institut, CH-5232 Villigen, Switzerland}
\author{E. Cappelli}
\affiliation{Department of Quantum Matter Physics, University of Geneva, 24 Quai Ernest-Ansermet, 1211 Geneva 4, Switzerland}
\author{S. Kumar Chaluvadi}
\affiliation{Istituto Officina dei Materiali (IOM)-CNR, Laboratorio TASC, Area Science Park, S.S. 14 km 163.5, Trieste I-34149, Italy}
\author{E. Bonini Guedes}
\affiliation{Swiss Light Source, Paul Scherrer Institut, CH-5232 Villigen, Switzerland}
\author{M. Naamneh}
\affiliation{Swiss Light Source, Paul Scherrer Institut, CH-5232 Villigen, Switzerland}
\author{G. Vinai}
\affiliation{Istituto Officina dei Materiali (IOM)-CNR, Laboratorio TASC, Area Science Park, S.S. 14 km 163.5, Trieste I-34149, Italy}
\author{J. Fujii}
\affiliation{Istituto Officina dei Materiali (IOM)-CNR, Laboratorio TASC, Area Science Park, S.S. 14 km 163.5, Trieste I-34149, Italy}
\author{P. Torelli}
\affiliation{Istituto Officina dei Materiali (IOM)-CNR, Laboratorio TASC, Area Science Park, S.S. 14 km 163.5, Trieste I-34149, Italy}
\author{I. Vobornik}
\affiliation{Istituto Officina dei Materiali (IOM)-CNR, Laboratorio TASC, Area Science Park, S.S. 14 km 163.5, Trieste I-34149, Italy}
\author{A. Goldoni}
\affiliation{Elettra Sincrotrone Trieste, s.s. 14 km 163.5 in Area Science Park, 34149 Trieste, Italy}
\author{P. Orgiani}
\affiliation{Istituto Officina dei Materiali (IOM)-CNR, Laboratorio TASC, Area Science Park, S.S. 14 km 163.5, Trieste I-34149, Italy}
\affiliation{CNR-SPIN, UOS Salerno, Fisciano, SA, I-84084, Italy}
\author{F. Baumberger}
\affiliation{Department of Quantum Matter Physics, University of Geneva, 24 Quai Ernest-Ansermet, 1211 Geneva 4, Switzerland}
\author{M. Radovic}
\affiliation{Swiss Light Source, Paul Scherrer Institut, CH-5232 Villigen, Switzerland}
\author{G. Panaccione}
\affiliation{Istituto Officina dei Materiali (IOM)-CNR, Laboratorio TASC, Area Science Park, S.S. 14 km 163.5, Trieste I-34149, Italy}%

\date{\today}

\begin{abstract}
\vo\ has long been studied as a prototypical strongly correlated material.
The difficulty in obtaining clean, well ordered surfaces, however, hindered the use of surface sensitive techniques to study its electronic structure.
Here we show by mean of X-ray diffraction and electrical transport that thin films prepared by pulsed laser deposition can reproduce the functionality of bulk \vo\.
The same films, when transferred \is, show an excellent surface quality as indicated by scanning tunnelling microscopy and low energy electron diffraction, representing a viable approach to study the metal-insulator transition (MIT) in \vo\ by means of angle-resolved photoemission spectroscopy.
Combined, these two aspects pave the way for the use of \vo\ thin films in device-oriented heterostructures.
\end{abstract}

\maketitle


\section{Introduction}

\begin{figure}
\centering
\includegraphics[width=\columnwidth]{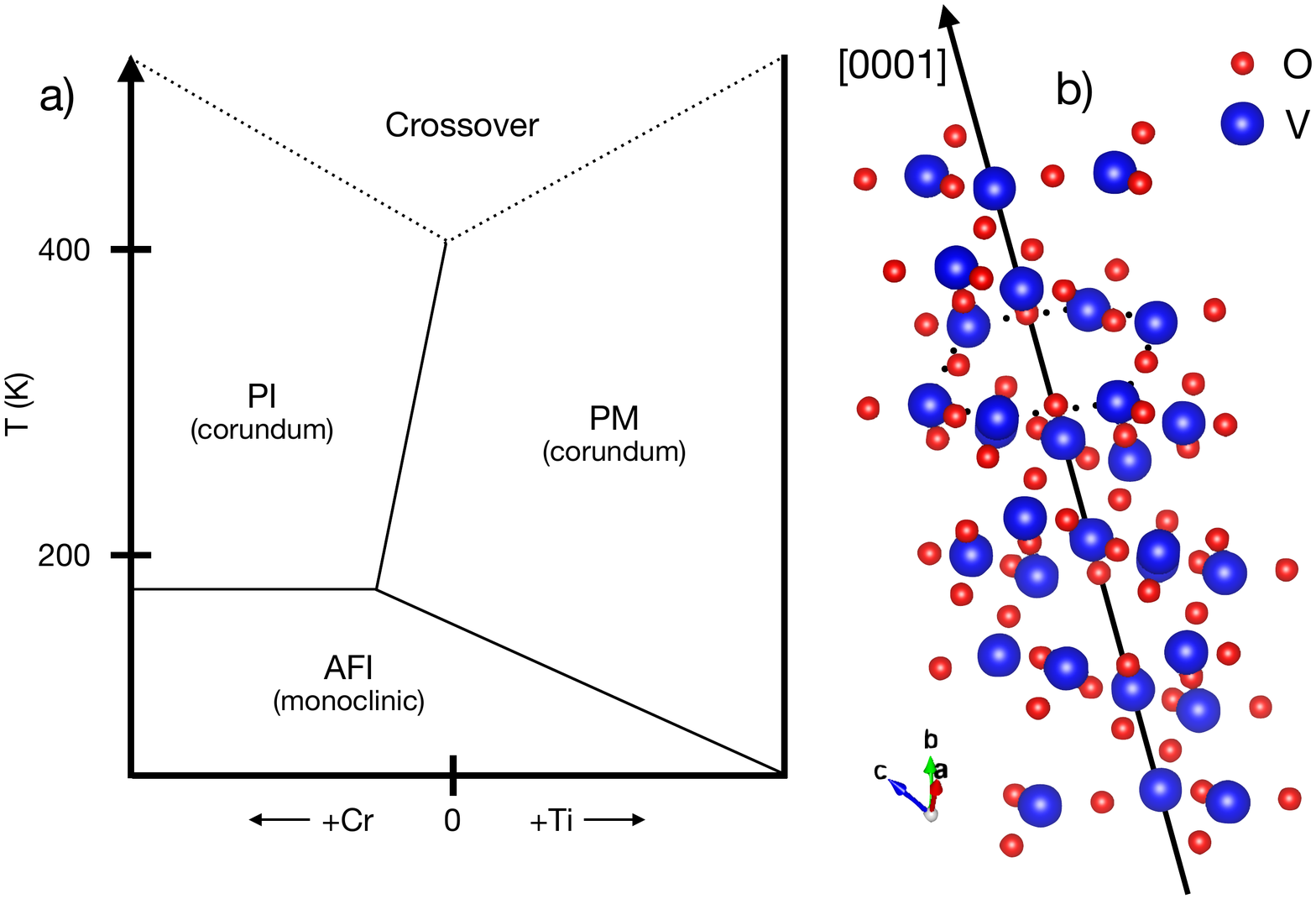}
\caption{(a) The temperature-doping phase diagram of \vo\. (b) \vo\ crystal structure in its high-temperature corundum phase: in blue (red) the V (O) atoms. The arrow indicates the [0001] direction, whose surface is the one investigated here, while the black dotted line is a guide for the eye to underline the honeycomb lattice surface perpendicular to the [0001] direction. }
\label{phasediagram}
\end{figure}

Since decades \vo\ and Cr-doped \vo\ are studied as prototypical systems for studying a pure Mott-Hubbard transition \cite{mott68}.
The temperature-doping phase diagram of \vo\ is presented in figure \ref{phasediagram}: undoped \vo\ shows a paramagnetic (PM) - antiferromagnetic insulator (AFI) electronic phase transition, accompanied with a corundum - monoclinic structural phase transition at temperatures around 150 K.
This temperature can be lowered using Ti doping, down to the point where it vanishes for doping levels greater than 5\%.
On the contrary, Cr doping brings the occurrence of a paramagnetic insulator (PI) phase, that shares the same corundum structure of the parent compound.

The trigonally distorted octahedral crystal field of vanadium sesquioxide splits the $t_{2g}$ states into a doubly degenerate $e_{g}^{\pi}$ and a nondegenerate $a_{1g}$ states, to be filled by the two valence electrons of V\textsuperscript{3+} atoms \cite{Ezhov99}.
Initial low spin - single band models used to describe the electronic structure of \vo\ had to be abandoned when a polarisation-dependent X-ray Absorption Spectroscopy (XAS) study showed that both $e_{g}^{\pi}$ and $a_{1g}$ states are populated in all the three phases, albeit with different occupation ratio among them \cite{park00,mossanek07}.
In the meanwhile, dynamical mean-field theory (DMFT) studies approached the multi-orbital problem, unraveling a strong many-body enhancement of the trigonal crystal field splitting that brings the $e_{g}^{\pi}$ states almost entirely below the Fermi level, leaving the Fermi surface of the PM phase to be composed predominantly by $a_{1g}$ states \cite{poteryaev07}.
However, recent theoretical studies based on full charge consistency reduced this strongly polarised scenario \cite{Grieger14,Shining14}.

Until recently, no Angle resolved photoemission spectroscopy (ARPES) studies were available to guide DMFT calculations.
The first ARPES data presented by Lo Vecchio et al. \cite{lovecchio16}, however, challenged the current narrative around the electronic structure of the PM phase, showing the presence of both electron and hole pockets, that is a clear signature of a considerable weight of $e_{g}^{\pi}$ states at the Fermi level.
The authors were able to obtain a highly ordered (0001) surface of \vo\ by annealing a polished single crystal in oxygen rich atmosphere (oxygen-annealing).
However, controlling this preparation step is not trivial, and slight variations in oxygen pressure or annealing temperature will result in surfaces being far different from the truncated \vo\ bulk \cite{SURNEV03}, like vanadyl-terminated, oxygen or vanadium rich terminations \cite{SCHOISWOHL04}, or, in some cases, with vanadium in a completely different oxidation state \cite{VANBILZEN15}.

Using alternative approaches, however, it is possible to obtain surfaces that can be a good approximation of a bulk truncation.
Fabrication of thin films using pulsed laser deposition (PLD) avoids any of the steps traditionally used in surface preparation (such as high-energy ion sputtering or annealing in vacuum or oxygen-rich atmosphere) that might lead to surfaces that are not representative of the bulk.
This approach has been used in the past, for instance, to control the surface termination and correct doping of YB\textsubscript{2}C\textsubscript{3}O\textsubscript{7} \cite{Sassa11}, making possible to visualise the expected ortho-II band folding otherwise not visible in cleaved single crystals.
Moreover, using thin films adds yet new possibilities to manipulate material properties, like the emergence of thermodynamically unfavored phases \cite{surnev01}, dimensionality control \cite{king14,caputo20}, substrate induced strain \cite{Dhaka15}, and hetero-structuring \cite{PLUMB17,Polewczyk21}.
Finally, the use of thin films ensures a higher degree of control in the MIT occurrence, mitigating the known fragility that bulk \vo\ crystals, like other materials showing structural transitions, exhibit upon repeated transition cycles.

Besides studies of fundamental interest, \vo\ thin films can be one of the building blocks of useful heterostructures for the emerging field of Mottronics \cite{tokura17}, where a precise control of bulk functionalities, as well as surface and interface properties, are mandatory for the fabrication of reliable devices.
Critical aspects in oxide functional heterostructure for device applications are stoichiometry (which defines the bulk functionalities), and surface/interface terminations (in order to modify the bulk functionality with the other blocks of the heterostructure).

In order to actually use \vo\ thin films as a reliable platform for fundamental studies, but also as a starting point for device-oriented heterostructures, finding a growth protocol that results in a functional \vo\ bulk, and that shows at the same time a surface representative of the bulk properties and functionalities, is mandatory.
Here we show that this goal can be achieved with thin films prepared by PLD.
X-ray diffraction and transport measurements indicate a bulk-like growth, with a clear MIT occurring around 185 K.
Surface quality has been verified by X-ray photoemission spectroscopy (XPS), XAS, scanning tunnelling microscopy (STM) and low energy electron diffraction (LEED), showing long range order and correct \vo\ surface electronic structure.
Moreover, comparing \is and \es transferred sample, spectroscopy indicates a clear surface degradation, stressing the importance for all the surface studies of an \is transfer protocol.
Finally, ARPES could capture the band reorganisation occurring at the MIT at the very topmost layers of \vo.


\section{Methods}

Samples were grown starting from commercially available \vo\ targets (SurfaceNet GmBH and Stanford Advanced Materials) ablated using the fundamental harmonic of a Nd:YAG laser (1064 nm) onto (0001)-oriented sapphire (Al\textsubscript{2}O\textsubscript{3}) substrates \cite{Chaluvadi21}.
Depositions were performed in vacuum ($p<10^{-7}$mbar), with the substrate kept at a temperature of 750$^\circ$C.

Electrical characterisation was carried out by standard four probes dc technique in the van-der-Pauw configuration, with a bias pulsed and reversed current of 1mA.
Scanning tunnelling microscopy (STM) experiment was performed with an atomic-resolution UHV apparatus, immediately after the growth of the samples.
STM topographic maps were acquired in a constant-current mode by using a tungsten tip.

XPS measurements were performed at the APE-HE beamline at the Elettra Synchrotron, using a Scienta R3000 hemispherical electron energy analyzer, with the sample at 45$^\circ$ with respect to the impinging linearly polarized light and the analyser normal to the surface.
The energy resolution for both XAS and XPS was set to 100 meV.
XAS measurements were taken in total electron yield (TEY) mode, normalizing the intensity of the sample current to the incident photon flux current at each energy value.
ARPES measurements were performed at the HighRes endstation of the SIS beamline at the SLS, using a Scienta R4000 electron alayser.
In this case, the total energy resolution was better than 20 meV.

Both systems host a PLD setup connected \is to the experimental chambers where synchrotron-radiation-based measurements are performed.

\section{results and discussion}

\subsection{Structural characterisation and transport properties}

\begin{figure*}
\centering
\includegraphics[width=1.8\columnwidth]{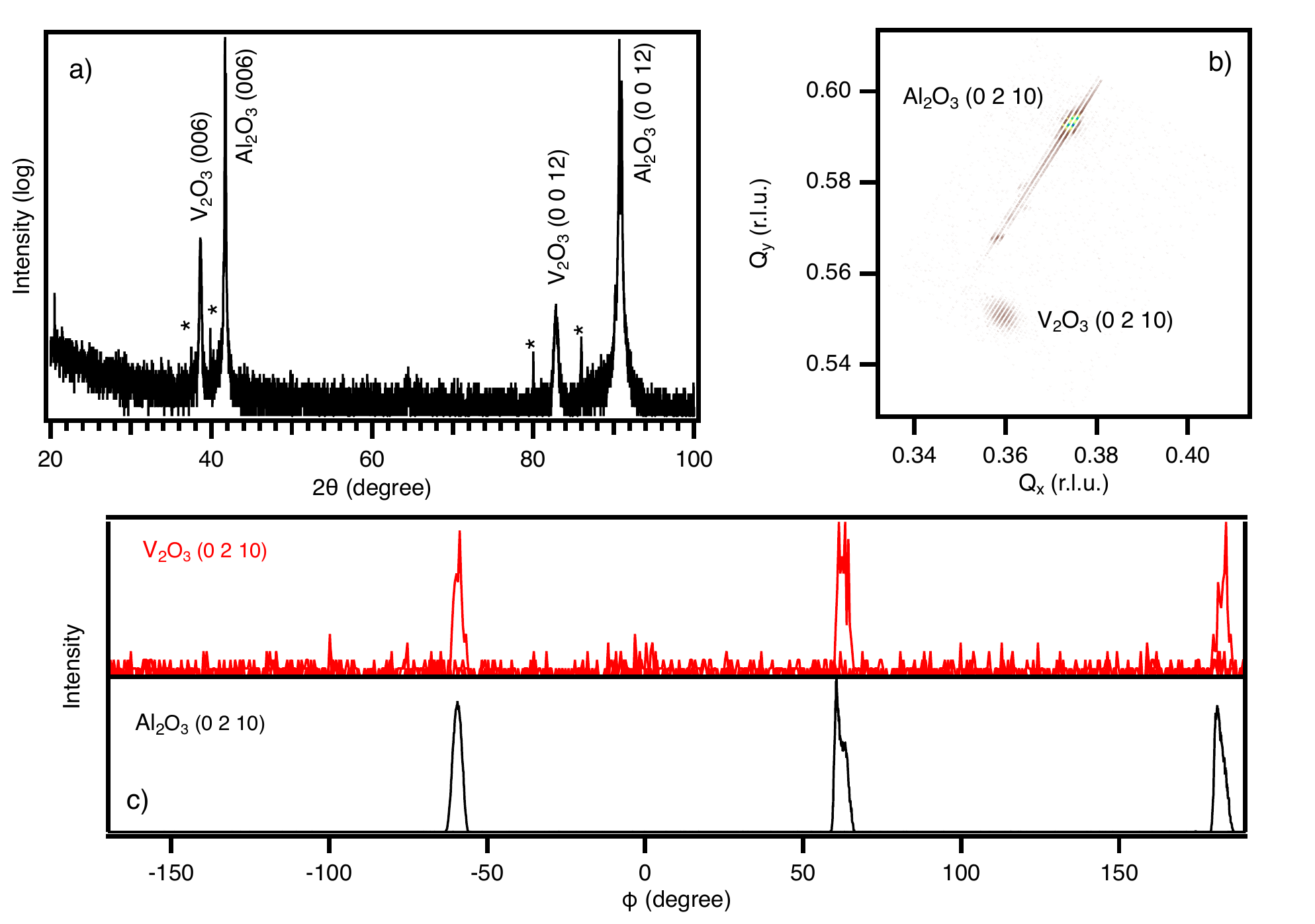}
\caption{(a) Symmetrical $\theta-2\theta$ scan of a \vo\ film grown on an \sap substrates. Peaks marked with a star are originated by the radiation emitted by the $K_{\beta}$ line of the Cu anode. (b) Reciprocal space map around the (0\,2\,10) asymmetric Bragg reflections of both film and substrate - Q\textsubscript{x} and Q\textsubscript{y} are reported in reciprocal lattice unit (r.l.u.); (c) Azimuthal $\phi$-scan of both film and substrate around the (0\,2\,10) asymmetric Bragg reflections.}
\label{xrd}
\end{figure*}

Structural characterisation of a 56 nm thick \vo\ films was performed by means of \es x-ray diffraction (XRD) and x-ray reflectivity (XRR), as shown in Figure \ref{xrd}.
XRD symmetrical $\theta-2\theta$ scans of \vo\ films only contain (0\,0\,$6l$) peaks, indicating the preferential c-axis orientation of the film along the [0001] crystallographic direction of the Al$_{2}$O$_{3}$ substrates and confirming the R-3c:H trigonal crystal structure.
Structural data was compared to those reported in literature for bulk \vo\ \cite{finger80}. In particular, \vo\ shows a R-3c trigonal space-group, with hexagonal cell lattice parameters a=4.9018 Å and c=13.9690 Å.
The out-of-plane parameter was determined by the (006) Bragg diffraction peak in the symmetrical $\theta-2\theta$ scan resulting in 13.97 $\pm$ 0.01 Å; in-plane parameter was determined by the (0 2 10) asymmetric Bragg reflection, which was expected at Q\textsubscript{x} and Q\textsubscript{y} values of 0.35739 and 0.55260 r.l.u., respectively; the experimental value of Q\textsubscript{x} was found at 0.358 r.l.u., corresponding to a=4.89 $\pm$ 0.02 Å; 
By considering the experimental values, the lattice parameters of our thin films fully match those reported for bulk \vo, therefore confirming that films are fully relaxed.
Interestingly, the azimuthal $\phi$-scan of \vo\ only shows a three-fold symmetry, as for the \sap substrate with no trace of twin-domain with a 60 degree in-plane rotation.

\begin{figure}
\centering
\includegraphics[width=0.9\columnwidth]{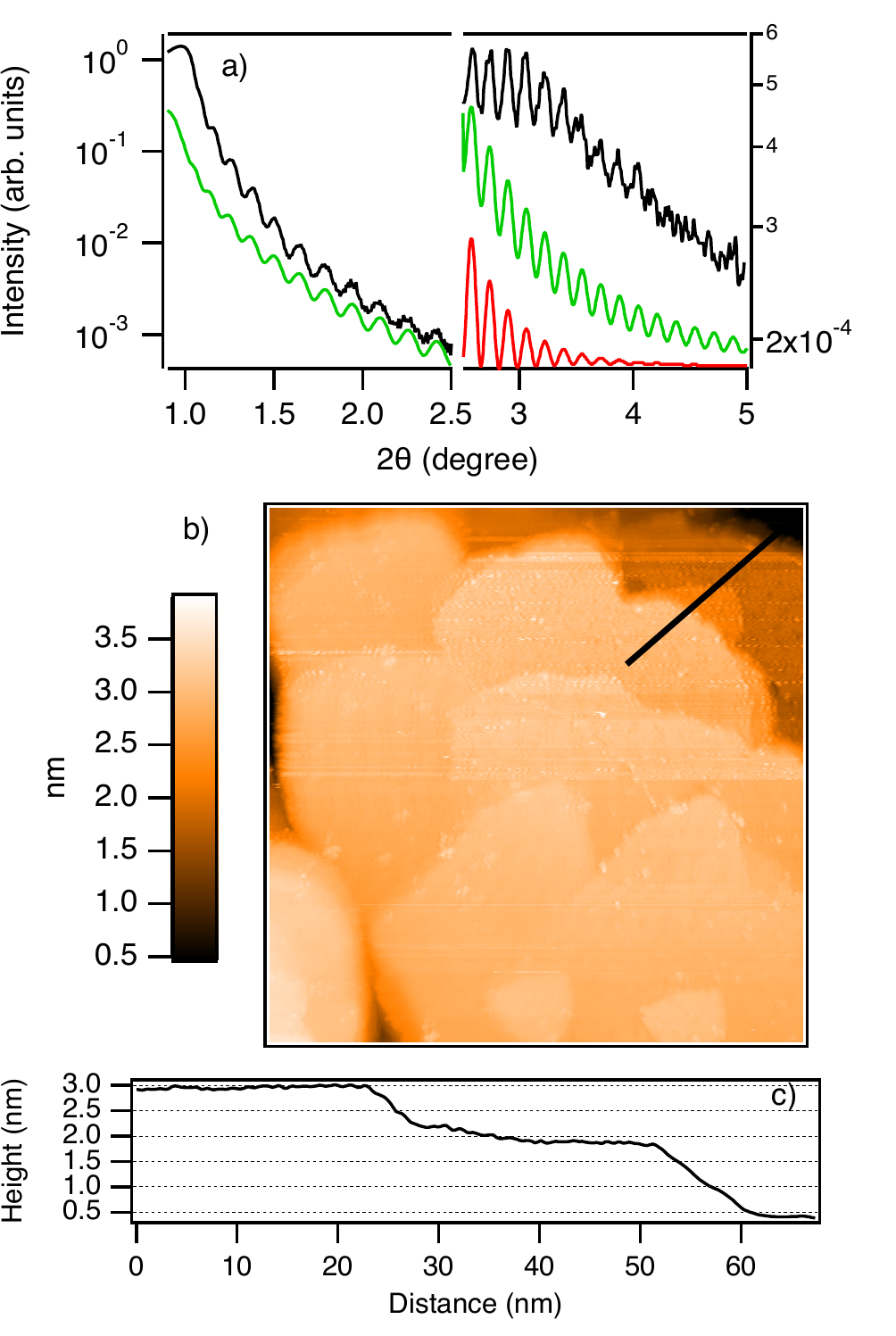}
\caption{(a) XRR curve of a \vo\ film grown on (0001)-\sap substrate: the black curve is the experimental data, compared to a simulated signal from an atomically perfect surface (green curve) and a simulated 1-nm rough surface (red curve). (b) Surface topography seen by STM, with a profile line (panel c) taken along the black line.}
\label{structure}
\end{figure}

Film thickness and surface roughness were investigated by low-angle X-ray reflectivity (XRR) (simulations of the data have been performed by means of the IMD package of XOP software), as shown in Figure \ref{structure}a.
Oscillation simulated for a film roughness of 1 nm (red curve) are clearly damped for $\theta-2\theta$ values greater than 4 degrees, while still visible on the experimental data (black curve), setting an upper limit for surface roughness.
In order to give a better estimation for the mean surface roughness we performed \is STM.
Figure \ref{structure}b shows a 150x150 nm image representative of the whole sample.
Large terraces of the order of tens of nanometers are visible, with steps height of 0.6 or 1.2 nm (panel c, line profile along the line in panel b).
The overall root mean square (RMS) height variation calculated over the same image is about 0.8 nm, indicating a very low surface roughness of the grown samples, consistent with the XRR data.

\begin{figure}
\centering
\includegraphics[width=0.9\columnwidth]{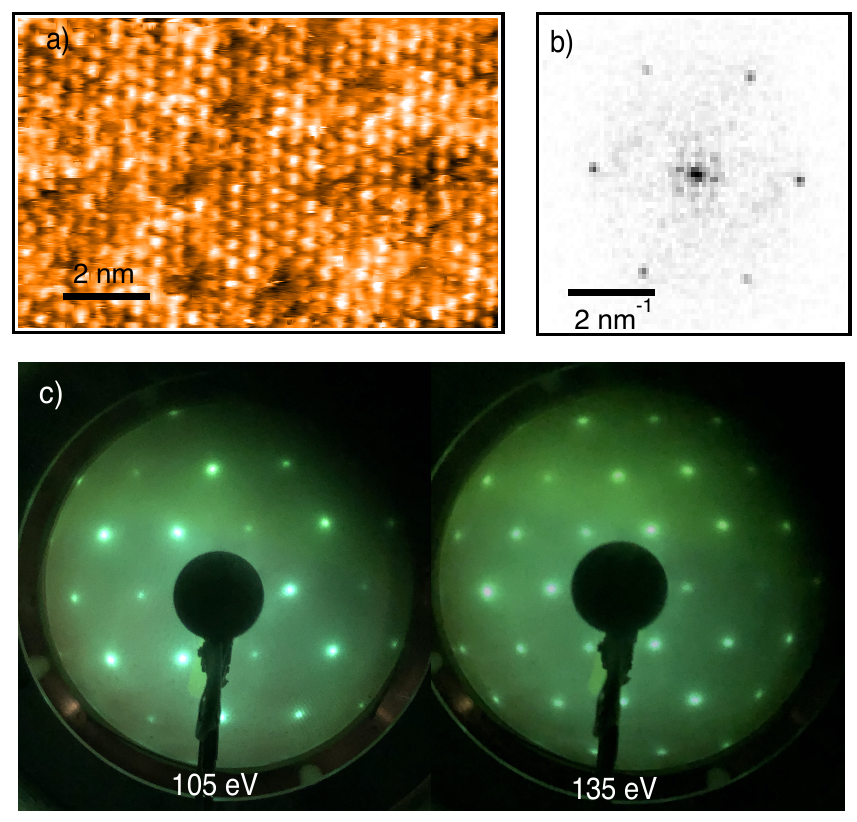}
\caption{(a) Atomic resolution STM image, with its 2D-FT (b). (c) LEED patterns of a \vo\ film acquired using electrons of 105 eV and 135 eV kinetic energy.}
\label{surface}
\end{figure}

Excellent surface quality is also confirmed by atomic resolution STM (figure \ref{surface}a), and its corresponding two dimensional Fourier transform (2D-FT), showing perfect hexagonal symmetry.
The long-range order of the surface was also probed by \is LEED.
Figure \ref{surface}b reports the LEED patterns obtained using primary electron with kinetic energy of 105\,eV and 135\,eV respectively.
The pattern shows sharp diffraction spots arranged with a hexagonal symmetry, without any surface reconstruction, and without circular rings connecting them, indicating no trace of randomly oriented domains.
By changing the primary energy it is possible to vary the intensity ratio between the two subsets of threefold-symmetric spots, indicating a threefold symmetry for the three-dimensional crystal and a growth with predominantly one domain orientation.

\begin{figure}
\centering
\includegraphics[width=0.9\columnwidth]{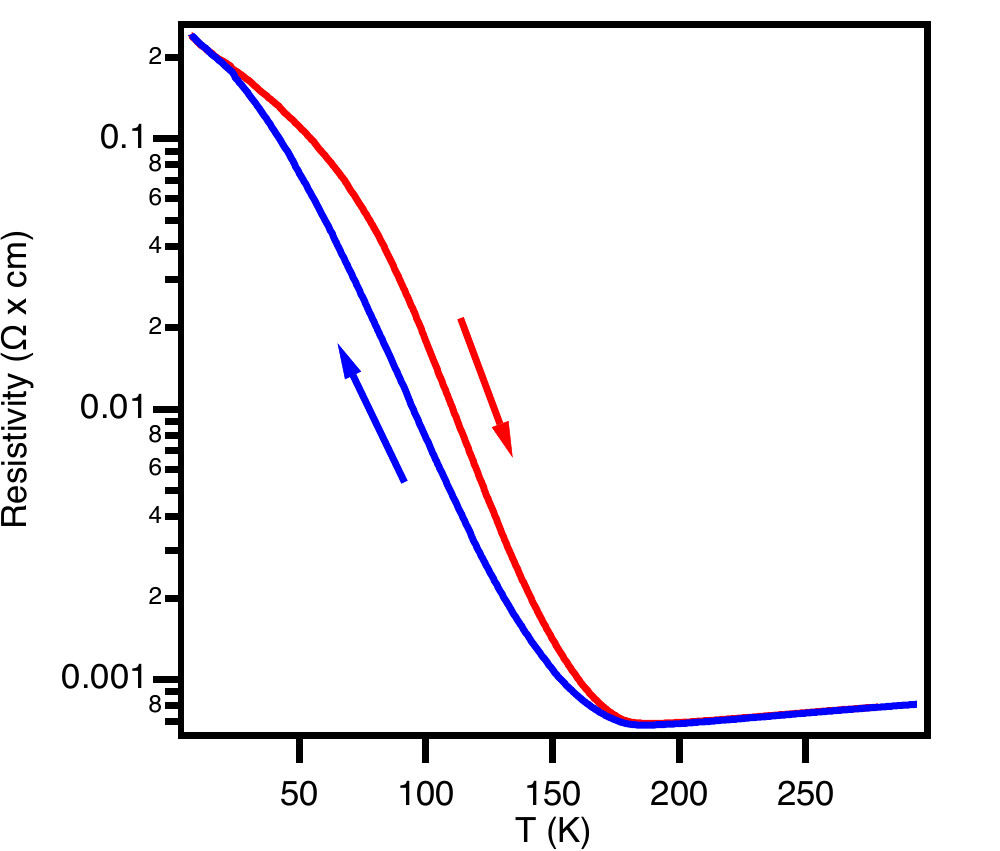}
\caption{Resistivity behaviour as a function of the temperature of a \vo\ thin film measured during a cooling (blue) / warming (red) cycle.}
\label{transport}
\end{figure}

Finally, transport properties of \vo\ (Figure \ref{transport}) show the hysteretical behaviour of resistivity as a function of temperature expected for its MIT.
It is worth noting that the thermal hysteresis of the electronic transport properties might not be as pronounced as in single crystals (where it is several orders of magnitude) because of the structural disorder characteristic of thin films.
In particular, as also observed in other oxide systems \cite{dagotto11}, structural disorder and inhomogeneities - characteristic of thin films - promote phase-separated domains, each of them characterised by a first-order MIT with slightly different transition temperatures.
This last can result into a \textit{broader} transition in resistivity measurements.

\subsection{Core level spectroscopy characterisation}

XAS and XPS experiments were performed on both \is and \es transferred \vo\ samples.
\textit{In-situ}-transferred samples were moved under UHV conditions (base pressure $< 2 \cdot 10^{-10}$\,mbar) to the end station of the APE-HE beamline, directly after the growth process.

\begin{figure}
\centering
\includegraphics[width=0.9\columnwidth]{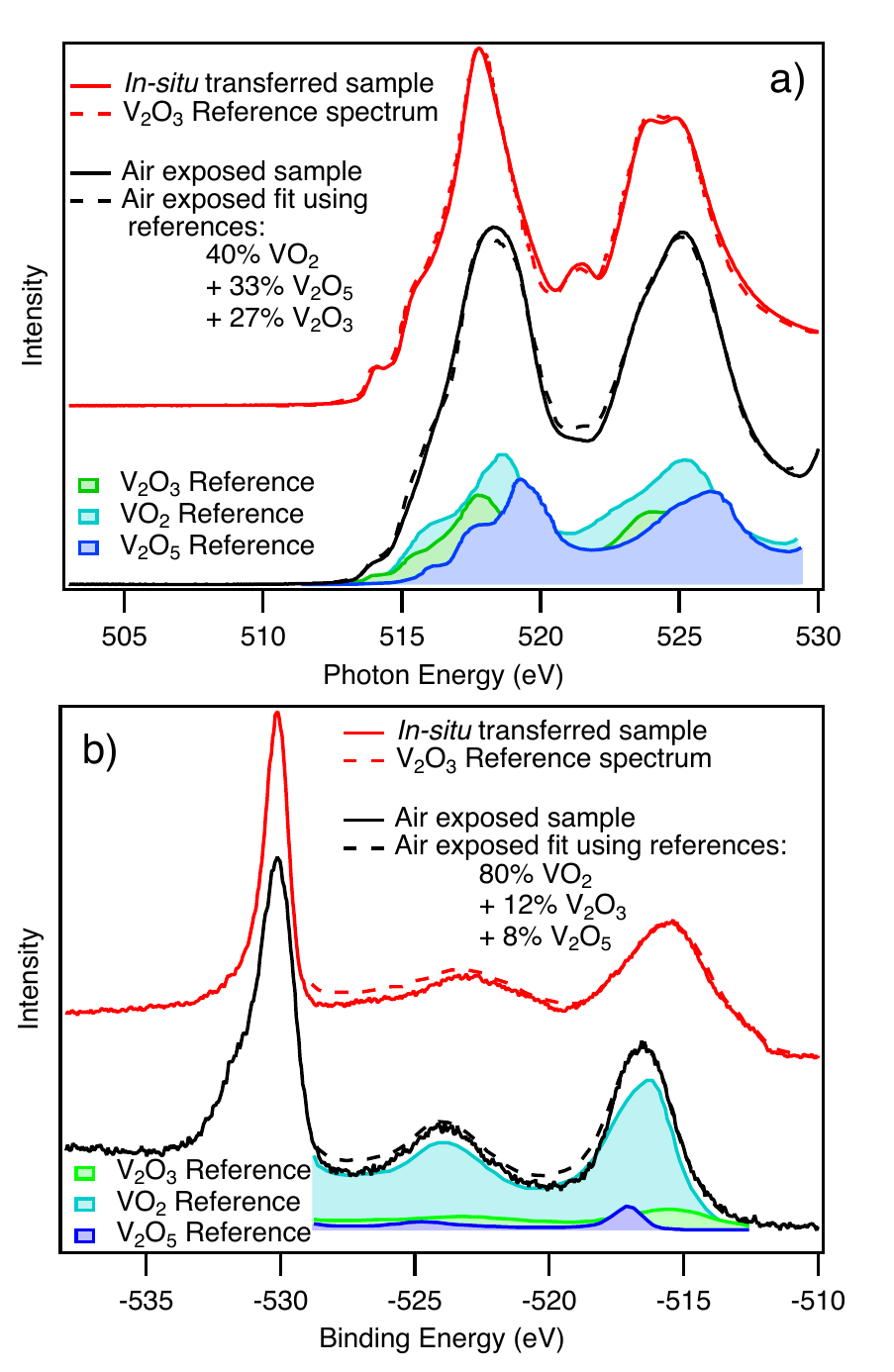}
\caption{a) XAS spectra at V L$_{2,3}$ absorption edges of \is (red) and \es (black) transferred \vo\ thick films; they are compared with reference spectra of \vo\, VO$_{2}$, and V$_{2}$O$_{5}$ form \cite{Zimmermann98}. b) V 2p and O 1s XPS peaks for the same V$_{2}$O$_{3}$ samples measured with a photon energy of 900\,eV and compared with cleaved single crystals reference spectra; references from \cite{Zimmermann98}.}
\label{corelevel}
\end{figure}

XAS spectra of the V L$_{2,3}$ absorption edges of \is transferred and \es \vo\ films are shown in figure \ref{corelevel}a.
Comparing the two, a substantial degradation of the \vo\ films upon exposure to air is recognisable.
The \is transferred sample (red curve) shows the same features measured on the reference \is cleaved single crystal sample (red dashed line, from \cite{Zimmermann98}), in particular a sharp L$_{3}$ edge at 517.7 eV, anticipated by a shoulder peak at 515 eV and a small pre-edge contribution at 514 eV, followed by a significant extra peak in the valley region between L$_{3}$ and L$_{2}$ peaks, with the latter presenting a quite marked double-peak feature.
All these detailed features are expected from both theoretical calculations \cite{ABBATE93,park00} and from previously reported experimental results on \is cleaved single crystals \cite{ABBATE93,muller97,Zimmermann98} and thin films \cite{sass03}.
On the other hand, the \es sample (black curve) presents much broader features, with a wider L$_{3}$ edge and no double peak on the L$_{2}$ edge: a clear sign of surface degradation.
The shift of the peaks and the marked change in shape of the L$_{2}$ edge could be tentatively attributed to a contribution coming from the formation of VO$_{2}$ and/or V$_{2}$O$_{5}$ \cite{ABBATE93} upon air exposure, or to a significative change in the electronic configuration of the topmost layers induced by molecular adsorption.
In order to investigate in detail this aspect, the XAS spectrum of the air exposed sample is compared with a linear combination of reference spectra acquired on \vo\, VO$_{2}$, and V$_{2}$O$_{5}$ single crystals (black dashed line, references from \cite{Zimmermann98}).
From this comparison, it is evident that just a small fraction (roughly one third) of the total XAS intensity can be attributed to \vo\, while the predominant contribution arises from vanadium in a different oxidation state (4+ for VO$_{2}$, and 5+ for V$_{2}$O$_{5}$).

Similarly, significant differences arise when comparing the V 2p XPS spectra of the \is and \es \vo\ (figure \ref{corelevel}b - $h\nu$=900 eV).
In particular, the \es sample does not show the shake-down shoulder at 512.5 eV binding energy.
This feature is a fingerprint of the metallic state of \vo, being interpreted as a more efficient core-hole screening \cite{panaccione06}, and its absence indicate the absence of metallicity of the topmost layers.
In this case the fraction of signal originated from \vo\ is even lower than the one observed by XAS.
This observation could be attributed to the higher surface sensitivity of photoemission with respect to XAS acquired in TEY mode.
The distinct fractions of \vo\ identified by XAS and XPS is thus consistent with a scenario in which the healthy \vo\ bulk is buried under layers of crystal degraded by the air exposure.

It is also worth noticing that the O1s core level (centred at 530 eV of binding energy) is strongly affected by exposure to air: a new shoulder at higher binding energy appears, most likely due to the appearence of hydroxyl/vanadyl groups at the surface \cite{SURNEV03}.

\subsection{ARPES investigation}

Lastly, we turn our attention to the possibility of studying the MIT using ARPES on the \is-transferred samples.
For this purpose we grew a \vo\ film and transferred it \is in the high resolution end station of the SIS beamline at the Swiss Light Source.

\begin{figure*}
\centering
\includegraphics[width=1.8\columnwidth]{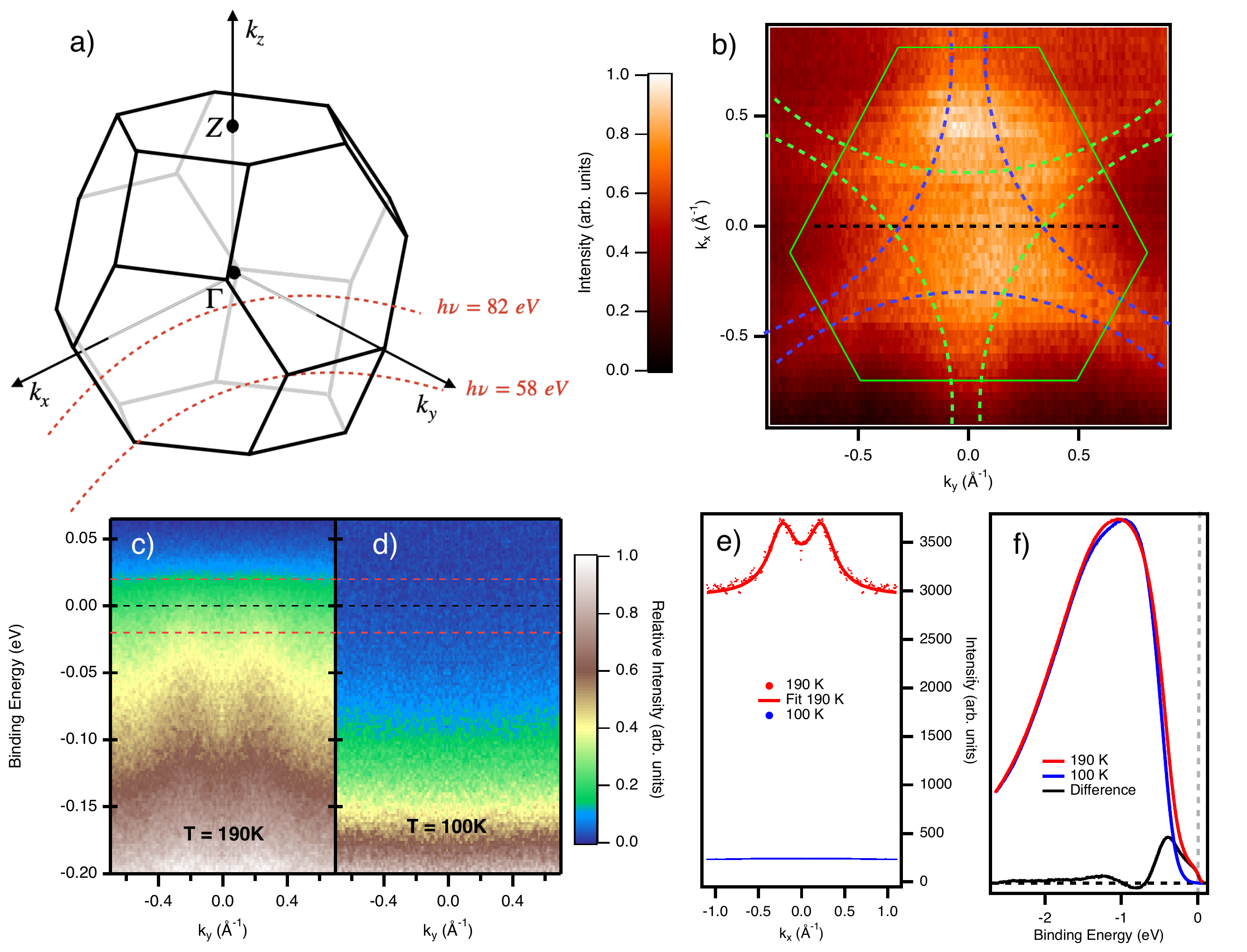}
\caption{Panel a: Brillouin Zone of \vo\ with indicated the $k_x$, $k_y$, and $k_z$ notation used in the following. Red dashed lines indicates the cuts in the $k_z$ dimension at the used photon energy. Panel b: Fermi surface acquired at a photon energy of 58 eV at 190 K. The black dashed line represents the position where the cuts shown in panels c and d are taken, while the blue and green dashed lines are guide to the eyes to mark the arc-like structures described in the text. Panels c and d: E\textsubscript{B} vs $k$ cuts acquired with a photon energy of 82 eV at 190 K and 100 K respectively. The black dashed lines mark the position of the Fermi level, while the red dashed lines denote the energy interval used to create the MDCs reported in panel e. Panel e: comparison between MDCs at low and high temperature. Panel f: comparison between the momenta integrated spectra in a wider range above (red) and below (blue) the MIT temperature, together with their difference spectrum (black).}
\label{arpes}
\end{figure*}

Figure \ref{arpes}b shows a cut in the $k_x$ - $k_y$ plane of the Fermi surface (in the following, for semplicity, just FS) of the \vo\ film acquired at a temperature of 190K with a photon energy of 58 eV.
This choice of photon energy allows to cut the three dimensional Brillouin zone (BZ) roughly half way between $\Gamma$ and Z (see figure \ref{arpes}a: red dashed lines indicate the cuts in the BZ corresponding to the two photon energies used in this study, and calculated using an inner potential value of 14 eV as in ref \cite{lovecchio16}).
Probing this region of the reciprocal space permits an easier comparison with the one already published in \cite{lovecchio16} (acquired on polished and oxygen-annealed single crystal).
The shape of the FS resembles the one measured in ref \cite{lovecchio16}, with arc-like structures that leave an hollow in the BZ center.
The most striking difference between the two is the symmetry of this structure: threefold in the case of ref \cite{lovecchio16}, while sixfold in our case.
Most likely this difference arises from the presence of two 60 degrees rotated domains in the film, resulting in an apparent sixfold symmetry instead of the intrinsic threefold one.
The two sets of dashed lines (green and blue) highlight the contribution coming from the two rotated domains, while the green continuous line marks the edge of the first Biroullin zone for the green-oriented domain.

In Figure \ref{arpes}c we report the (symmetrised) E\textsubscript{B} vs $k$ cut along the dashed line shown in panel b, but acquired with a photon energy close to the bulk $\Gamma$ point (82 eV), in which a faint parabolic-shaped feature emerges from the incoherent background.
A momentum-distribution curve (MDC) integrated across the Fermi level among the red dashed lines (reported in panel e, red dots) can be fitted using two Lorentzian functions (red continuous line) $I=\frac{1}{\pi}\frac{\Gamma}{(x-x_0)^2+\Gamma^2}$, with the Fermi momentuim $k_F=x_0$ and the electron mean free path $\lambda=\frac{1}{\Gamma}$.
The Fermi momentum extracted from our fit is $k_F=\pm0.23\,\AA^{-1}$, in fair accordance with those find in ref \cite{lovecchio16}.
The energy distribution-curve (EDC) integrated over all the momenta (panel f, red) shows a clear Fermi cut.

Upon cooling down the sample to 100K, the parabolic feature, as well as the incoherent spectral weight present at the Fermi level, vanishes, as clearly visible from Figure \ref{arpes}d, and its EDC and MDC reported in panels e and f.
The comparison between the MDCs at low and high temperature, reported in Figure \ref{arpes}e, shows a clear suppression of spectral weight at the Fermi level.
Finally, the angle-integrated spectra at high (red) and low (blue) temperature are reported in figure \ref{arpes}f, together with their difference spectrum (black).
The latter shows a spectral weight redistribution from lower to higher E\textsubscript{B} in the range from -0.7 to -1.5 eV.

It is worth noticing that the whole spectral weight between the Fermi level and 0.2 eV, i.e. where the quasiparticle lies, vanishes as a result of the metal-to-insulator transition taking place.
This behaviour is similar to the one observed in nickelates, where it has been proposed that a charge disproportionation driven by temperature decrease \cite{johnston14} or dimensionality effect \cite{king14} results in increasing charge localisation and hence formation of a large gap in the DOS \cite{Dhaka15}.

\section{Conclusion}

Ancillary characterisation showed how our PLD-grown films grow atomically smooth, without any sign of surface reconstruction.
A clear MIT could be detected using a four-probe transport system, even though the intrinsic disorder of thin films, if compared to single crystals, widens its temperature range.
The use of core level spectroscopy pointed out the importance of \is transfer, in order to preserve surface properness, fundamental in ARPES studies.

Finally, a broad dispersive band could be observed in ARPES.
Upon cooling, the MIT manifested itself with the complete disappearance of this band.

More extensive ARPES investigation of these films would be highly desirable, possibly with the use of photon energies not limited in the VUV range, but also extending to the more bulk sensitive regimes of the soft and hard X-rays.
This is motivated by the suggestion of different electronic correlations at bulk and surface regions of the sample \cite{borghi09,panaccione06,rodolakis09,Papalazarou09}, which has generally been termed ``surface dead layer'' in many transition-metal oxides, as e.g. manganites and titanates.
The use of \is transfer of samples, as in the present study, allows a better control of the surface environment, and may help to address this open question.
The same holds true also for time- and space-resolved ARPES studies, that in the recent past shed new light on the MIT in pristine and doped vanadium sesquioxide \cite{lantz17,lupi10}.
Moreover, the thin film approach will also allow to explore new directions in the study of the archetypal model system \vo\: strain and thickness are two parameters that can easily be adjusted during the thin-film growth and that could result in a tuning of the MIT \cite{Dillemans14,Thorsteinsson18}.
Additionally, strain could be used to decouple the electronic and structural transition, allowing the study of their individual roles \cite{Majid17}.
Proximity effects with functional layers can also induce modifications in the properties of vanadium sesquioxide, by, for instance, controlling the magnetic frustration and long range AFM ordering \cite{Leiner19,Trastoy20}, or controlling the strain by using piezoelectric substrates \cite{Sakai19}.

\section{Acknowledgment}
This work has been partially performed in the framework of the Nanoscience Foundry and Fine Analysis (NFFA) project. We thank Elettra for providing beamtime under the proposal number 20190319.


\bibliography{apssamp}

\end{document}